\documentclass[twoside,slac_one]{revtex4}
\usepackage{graphicx}
\usepackage{fancyhdr}
\usepackage{amsmath} 
\usepackage{bm}
\usepackage{amsxtra}
\usepackage{amssymb}
\usepackage{amsthm}
\usepackage{latexsym}
\usepackage{lscape}

\begin{document}

\title{Search for the Neutron Electric Dipole Moment: Contributions from the Triangle Universities Nuclear Laboratory }

%

\author{P. R. Huffman, R. Golub, C. R Gould, D. G. Haase, D. P. Kendellen, E. Korobkina, C. M. 
Swank, A. R. Young}
\affiliation{Department of Physics, North Carolina State University, Raleigh, NC, USA}
\author{M. W. Ahmed, M. Busch, H. Gao, Y. Zhang, W. Z. Zheng}
\affiliation{Department of Physics, Duke University, Durham, NC, USA}
\author{Q. Ye}
\affiliation{Oak Ridge National Laboratory, Oak Ridge, TN, USA}
\affiliation{National Institute of Standards and Technology, Gaithersburg, MD, USA}

\begin{abstract}
A significant fraction of the research effort at the Triangle Universities Nuclear Laboratory (TUNL) focuses on weak
interaction studies and searches for physics beyond the Standard Model.  One major effort is the development of a new
experimental technique to search for the neutron electric dipole moment (nEDM) that offers the potential for a factor of
100 increase in sensitivity over existing measurements.  The search for this moment has the potential to reveal new
sources of time reversal (T) and charge-conjugation-and-parity (CP) violation and to challenge calculations that propose
extensions to the Standard Model.  We provide a brief overview of the experiment as a whole and discuss the work
underway at TUNL as part of this effort.
\end{abstract}

\maketitle

\thispagestyle{fancy}


\section{Introduction}

The Golub-Lamoreaux proposal\cite{Gol94} to measure the neutron electric dipole moment (nEDM) using ultracold neutrons
(UCN) and polarized $^{3}$He in a superfluid bath of helium is being developed by the nEDM Collaboration, a consortium
of 21 institutions\cite{EDM}, and will be operated at the Spallation Neutron Source at Oak Ridge National Laboratory in
Oak Ridge, TN. The Triangle Universities Nuclear Laboratory (TUNL) is playing a major role in the development of many
key aspects of this experiment.

The experimental search for a neutron electric dipole moment has the potential to reveal new sources of time-reversal
(T) and charge-conjugation-and-parity (CP) violation and to challenge calculations that propose extensions to the
Standard Model.  The goal of the experiment is to improve the measurement sensitivity of the nEDM by two orders of
magnitude over the present experimental limit \cite{Bak06}.  A successful measurement will impact our understanding of
the physics of both weak and strong interactions.  The physics goals of this experiment remain timely and are of
significant importance \cite{Dub11}.

The experiment is based on a magnetic-resonance technique: a magnetic dipole is rotated in a magnetic field.  Polarized
neutrons and polarized $^{3}$He atoms are dissolved in a bath of superfluid $^{4}$He at a temperature of $< 450$~mK.
When placed in an external magnetic field, both the neutron and $^{3}$He magnetic dipoles precess in the plane
perpendicular to the magnetic field.  The measurement of the neutron EDM involves measuring the difference in the
precession frequencies of the neutrons and the $^{3}$He atoms when a strong electric field is applied either parallel or
anti-parallel to the magnetic field.  In this comparison measurement, the neutral $^{3}$He atom is assumed to have a
negligible electric dipole moment and serves as a volume co-magnetometer.

In principle, this new type of nEDM experiment could achieve more than two orders of magnitude improvement in the
experimental limit for the neutron EDM. This factor results from the possibility of an increased electric field due to
the excellent dielectric properties of superfluid $^{4}$He, an increase in the total number of ultracold neutrons
stored, and an increased storage time for these UCNs.  The current experimental nEDM bound ($d_{n} < 3 \times 10^{-26}$
$e\cdot$cm)\cite{Bak06}, however, is limited by magnetic field systematics, particularly known as the geometric phase
effect\cite{Har99,Pen04}.  The use of $^{3}$He as a volume co-magnetometer is crucial to the minimization of this
magnetic-field systematic\cite{Lam05,Bar06}.

TUNL plays an active role in all aspects of the nEDM experiment, with our goals of addressing outstanding issues related
to maximizing the sensitivity of the measurement, designing the experiment itself, and serving in key management
positions.  We are investigating the geometric phase effect, developing the data acquisition and slow controls system,
designing the cryogenic environment of the experiment, simulating the spin transport of polarized $^{3}$He, simulating
the transport of neutrons, and developing an apparatus to study systematic effects simultaneously using polarized
$^{3}$He and ultracold neutrons.  Brief discussions of these activities are presented below.

\section{Geometric Phase Effects}

In the large-scale neutron electric dipole moment apparatus, the interaction of the field $\vec{B}_{\mbox{\textit{eff}}}
\sim \vec{v} \times \vec{E}$ with magnetic field gradients will produce a frequency shift linearly proportional to the
electric field.  Known as the geometric phase effect and mimicking a true nEDM, this effect has emerged as a primary
systematic error limiting the precision of the next generation of nEDM searches.

In the context of nEDM experiments, this effect was first investigated experimentally and theoretically by Pendelbury
\textit{et al.}\cite{Pen04} and later by Lamoreaux and Golub\cite{Lam05}.  In a published work\cite{Bar06}, we
introduced an analytic form for the correlation function that determines the behavior of the frequency shift, and showed
in detail how it depends on the operating conditions of the experiment.  Using an analytic function for the case of gas
collisions, we averaged over a Maxwellian velocity distribution to calculate the temperature dependence of the frequency
shift for $^3$He diffusing in superfluid $^4$He.  Solutions to the phase shift and relaxation rates for fields with
arbitrary spatial dependence have recently been found for 2 and 3 dimensions valid for all mean free paths.

The results indicate that it may be possible to fine-tune the effect to a high degree by an appropriate choice of
operating temperature.  From an experimental point of view it is very appealing to try to make use of the zero crossings
to reduce the systematic effects to zero.  Our theoretical work indicates it is plausible to do this for $^3$He, by
tuning the temperature of the apparatus to take advantage of the $1/T^7$ dependence of the diffusion constant for $^3$He
in superfluid $^4$He \cite{Lam02}.  Specifically for the nEDM experiment, we have used our simulation results in
combination with $^{3}$He transport calculations to set the nominal operating temperature for the nEDM experiment of $<
450$~mK.

Experimentally, we have constructed an apparatus where polarized $^{3}$He can be injected into a deuterated rectangular
polystyrene cell filled with superfluid $^{4}$He at $T \leq 500$~mK. Initial experiments were aimed at verifying the
predicted depolarization rates of polarized $^{3}$He within a measurement cell.  We measured a depolarization
probability per bounce of $\sim 1 \times 10^{-7}$ using NMR excitation and detection coils and superconducting
constant-field and gradient coils\cite{Ye08,Ye09}.  

Since these initial measurements, we carried out a series of measurements of the appropriate correlation function at
different densities.  To compensate for the fact that the relaxation takes place at different frequencies, we ramped the
magnetic field so that measurements could be made at a constant frequency.  It is necessary to ramp the gradient on and
off to enable measurements of the polarization as a function of decay time.

The geometric-phase-effect measurements rely on theoretical calculations that show one can determine the effect by
measuring the $T_{1}$ relaxation rate of the $^{3}$He polarization in a magnetic field gradient.  Thus it is not
required to measure directly the ($\vec{v} \times \vec{B}$)-induced frequency shift.  If one applies a uniform magnetic
field gradient $\partial B_{z}/\partial z$ large enough so that it dominates all other field gradients, the relaxation
time $T_{1}$ can be determined by traditional NMR techniques, assuming wall relaxation can be neglected.  This technique
does not require an electric field and thus significantly simplifies the experiment.  We have taken data on the
geometric phase effect and are presently in the process of its analysis.

\section{Studies of Systematic Effects}

The nEDM project has identified a new opportunity now available that is aimed to significantly reduce the risk to the
project while shortening the time to a physics measurement.  The PULSTAR UCN source, located on the campus of NC State
University, will soon come online and become a world-class source of UCNs\cite{Kor07}, comparable in intensity to
existing sources at both the Institut Laue Langevin (ILL)\cite{Ste89} and Los Alamos National Laboratory
(LANL)\cite{Sau04}.  This source, coupled with the existing polarized $^{3}$He capabilities at TUNL, will enable
development of an ideal setup for investigating many UCN--$^{3}$He interactions at cryogenic temperatures.

The apparatus is being developed with the primary goal of exploring key systematic effects for the nEDM experiment as
well as performing characterization tests on a full-size measurement cell for the main nEDM apparatus.  The design goal
is to develop an apparatus that will allow tests of five key scientific areas outlined below to be performed in a single
actual size nEDM measurement cell filled with liquid helium residing in a uniform magnetic field (without an electric
field).  

The basic idea of the apparatus is that a measurement cell filled with liquid $^{4}$He at a temperature of $T < 450$~mK
will be placed within a uniform magnetic field.  Cryogenic conventional and superconducting magnetic shields will
surround this geometry, as well as thermal radiation shields and finally a series of additional external conventional
magnetic shields.

Several components must couple with the measurement cell in order to cool the sample, fill the cell with UCN, fill and
remove the polarized $^{3}$He, and detect the scintillation light from neutron capture.  First, the cell will be
thermally anchored to the mixing chamber of a $^{3}$He-$^{4}$He dilution refrigerator to provide cooling.  A fill line
for the liquid helium will be thermally anchored to the refrigerator and allow helium gas to be condensed within the
cell through a small capillary.  A second line will extend from above the cryostat to the cell to allow polarized
$^{3}$He to be introduced within the cell.  At the end of a measurement cycle, the unpolarized $^{3}$He must then be
removed from the liquid.  UCN from the PULSTAR source will be transported to the cell through a series of guides and
introduced into the cell through a valve similar in design to the actual nEDM valve.  Light from the neutron capture
scintillations will be wavelength shifted to visible wavelengths using deuterated fluors, transported away from the cell
through acrylic lightguides, and detected using photomultiplier tubes.  Pulses will be digitized and analyzed offline.
    
We have identified five key scientific areas where studies at TUNL can provide significant advances to the development
and implementation of the nEDM project:

\mbox{}\newline\noindent\textbf{Measurement of Scintillations Due to the Relative UCN--$^\textbf{3}$He Precession and
Demonstration of the Critical Dressing Technique}

The fact that the magnetic moments of the neutrons and $^{3}$He are only 10~\% different reduces the sensitivity to
background fields by an order of magnitude.  If the moments were exactly equal, there would be no effect at all.
Unfortunately, we have no direct control over the physics responsible for the observed magnetic moments; however, these
moments can be artificially modified by using `dressed spin' techniques.  In the presence of a strong oscillating
magnetic field, the magnetic moment will be modified, or `dressed', yielding an effective gyromagnetic ratio.  With
proper choice of magnetic fields, the magnetic moments of two species can be made equal.

In practice, the oscillating field is at right angles to the static field $B_{0}$ around which the spins are precessing.
In the absence of the oscillating field, one would see an oscillation in the scintillation rate arising from neutron
capture on $^{3}$He at the difference in the precession frequencies.  When the RF dressing field is applied, the
effective magnetic moments become modified, and at the condition referred to as `critical dressing', the two will have
the same precession frequencies.

If the nEDM is non-zero, the neutron precession frequency will be shifted by an amount $2d_{n}Ej_{0}(\gamma _{n}x)$,
where $d_{n}$ is the nEDM, $E$ the electric field, and $j$ is a Bessel Function.  Thus, the value of $x=x_{c}$ that
yields `critical dressing', or equal precession of the frequencies, is changed.  One thus measures the value of $x_{c}$
versus the electric field direction to extract a neutron EDM signal.

Estimates show that one can obtain densities of both UCN and polarized $^{3}$He in the measurement cell that are
comparable to those that will be obtained in the main experiment.  Thus this apparatus will allow us to fully test the
spin dressing technique using both polarized UCN and polarized $^{3}$He simultaneously in a measurement cell.  The
development of the necessary electronics and operating procedures will represent a substantial contribution to the main
experiment and is substantially expected to reduce the time to data collection.  In addition, the proposed feedback
system will remove the effects of the pseudo-magnetic field.

\mbox{}\newline\noindent\textbf{Measurement of the `Trajectory Correlation Function' for Systematic Error Quantification} 

Relaxation effects ($T_{1}$ and $T_{2}$) and the so-called geometric phase systematic error all depend on the trajectory
correlation functions of the $^{3}$He and UCN. These correlation functions form a family related by differentiation (or
integration) so measurement of one serves to determine the whole family.

While theoretically one does not expect the wall collisions to have a large impact on the results, this will be tested
empirically.  Unexpected effects like rather long residence times of the particles in surface pores for instance, could
affect the results, so it is important to have an independent measure of these correlation functions.

\mbox{}\newline\noindent\textbf{Detection of the $^\textbf{3}$He Pseudomagnetic Field in Search of Possible Double Resonance Effects}

The nEDM measurement can be influenced by the `pseudo-magnetic field' discussed above.  This field arises from the
spin-dependent coherent scattering cross-section, which leads to an energy shift that is spin dependent and thus appears
as a magnetic field.  The pseudomagnetic field is not directly affected by the application of an electric field, but can
be the source of precession frequency fluctuations and hence extra noise in the system.  The influence of the
pseudomagnetic field can be reduced by ensuring that the magnetometer spins have no component along the static magnetic
field, which is possible by careful control of the spin flip pulses.

Such pseudomagnetic fields have appeared in other EDM experiments, for example, in a $^{129}$Xe experiment where the
field was of order 1~mHz due to the presence of spin polarized rubidium that was used to polarize and detect the
$^{129}$Xe spin precession\cite{Vol84}.  This frequency, as an EDM in a 5~kV/cm field that was used in the experiment,
corresponds to $10^{-22}~e\cdot$cm, while the final experiment sensitivity is in the $10^{-26}~e\cdot$cm range.  This
level of discrimination results simply from the fact that the electric field does not directly affect the pseudomagnetic
field, and the spin of the rubidium was approximately orthogonal to the applied static magnetic field.

\mbox{}\newline\noindent\textbf{Development of Techniques for NMR Imaging of $^\textbf{3}$He} 

The idea of the $^{3}$He co-magnetometer is that it should sample the magnetic field in the same way that the field is
seen by the UCN, i.e. both species should have a uniform density.  However the density distribution of the $^{3}$He can
be distorted by heat currents flowing in the measurement cell.  The phonons constituting the heat currents act as a wind
on the $^{3}$He atoms and cause them to move with the heat current.  Diffusion against this current results in an
equilibrium inhomogeneous density distribution.

Standard NMR techniques can be used to study these effects.  Although our NMR frequency is quite low, it has been shown
that with reasonable gradients, resolutions on the order of 1~cm can be achieved, sufficient for our measurement cells.
Note that these measurements can be carried out without UCN, so they can be performed with a considerably higher
$^{3}$He density than that in the nEDM search.

\mbox{}\newline\noindent\textbf{Study Techniques for Reversing $\sigma _{^\textbf{3}\mbox{\tiny He \normalsize}}$, $\sigma _{\mbox{\tiny UCN
\normalsize}}$ and $B_{0}$ and Other Ancillary Measurements}

Establishing the experimental parameters for reversing the spins and magnetic fields will take a considerable length of
time during the commissioning phase of the apparatus.  Developing these techniques at PULSTAR will provide an
environment where one can perform a more complete study of these reversals without the external time constraints imposed
by the project schedule.
   
The apparatus can also be used for additional tests on actual measurement cells that could be installed in the main
apparatus.  These item include for example UCN and/or $^{3}$He storage and depolarization studies to optimize
measurement cell materials and fabrication techniques as well serving as a test bed for new cell designs and/or trouble
shooting cells that don't work in main apparatus.

\section{$^{3}$He Spin Transport and Injection into Liquid Helium}

Polarized $^{3}$He is used as a co-magnetometer in the nEDM experiment.  This helium, produced using an atomic beam
source, is infused into a bath of $\sim 400$~mK liquid helium that resides above the measurement cells.  The transport
of $^3$He spins through magnetic field gradients has been simulated in order to demonstrate whether nearly 100~\%
polarized $^3$He can maintain its polarization after injection from the atomic-beam source (ABS) into the collection
volume in the nEDM experiment.  In addition, a test of the $^4$He-film burner is envisioned in order to demonstrate
whether the film burner can maintain low $^4$He vapor pressure in the injection line.  This is crucial for $^3$He to be
successfully injected into the collection volume.  

\mbox{}\newline\noindent\textbf{$^3$He Depolarization Simulations}

Simulations have been performed to calculate how the spin precesses as the $^3$He atoms travel inside the transport tube
from the ABS exit interface to the collection volume.  The 4$^{th}$ order Runge-Kutta method was used to solve the Bloch
equation, which describes a spin $\vec{M}$ precessing in a magnetic field $\vec{B}$.  The vector form of the Bloch
equation is written as $\frac{d\vec{M}(t)}{dt}=\gamma \vec{M}(t)\times\vec{B}(\vec{r})$ where $\vec{r}$ is the position
vector of the spin.  At $t=0$, the spins are at the ABS exit interface, and at $t=T$ the spins have traveled to the
collection volume.  By integrating the Bloch equation over this timeframe, one can obtain the magnetization $\vec{M}(T)$
of spins inside the collection volume.  Because the magnetic field inside the collection volume is in the $\hat{x}$
direction, the $\hat{x}$ component of the spin, i.e. $M_x$, gives the polarization.

Different magnetic-field distributions $\vec{B}(\vec{r})$ give different final $^3$He polarizations in the collection
volume.  The goal of the simulation was to assist in the design of the magnetic field so that the final polarization of
$^3$He is at least 95~\%.  The latest magnetic-field design (March 2010 from Arizona State University) has two current
configurations: one with all the magnetic-field currents in the positive direction, so that the $B$ field in the
collection volume is in the $+\hat{x}$ direction, and the other with all the currents reversed, so that the $B$ field is
in the $-\hat{x}$ direction.  Nearly 10,000 events have been simulated for each configuration, giving a statistical
uncertainty of $\sim 1~\%$.  The polarization of $^3$He as a function of position was calculated.  Two regions leading
to polarization loss were identified.  One is at the ABS exit and is due to the strong and irregular fringe fields of the
ABS; the other is close to the collection volume, where the spins start to rotate from the $^3$He injection direction to
the $\hat{x}$ direction in order to match the holding field inside the collection volume.  The final polarization in the
collection volume is calculated to be 76~\% for the positive current configuration, and 79~\% for the negative current
configuration.

The nEDM experiment requires $>95$~\% polarization of $^3$He in the collection volume, thus improvements in the magnetic
field design are required.  The polarization loss at the exit of the ABS can be reduced by increasing the
transport-magnetic-field strength so that spins can follow field changes more easily.  Meanwhile, the spin rotation
should also be carried out at high fields near the ABS exit, instead of at low fields, which cause polarization loss.
Once spin rotation is complete, the transport field needs to be tapered down from a few hundred gauss to several hundred
milligauss to match the holding field in the collection volume.  Presently, the University of Kentucky group is working
on a new magnetic-field design that incorporates the features mentioned above.  

\mbox{}\newline\noindent\textbf{$^3$He Injection/Film Burner Tests}

The injection test was designed to determine whether polarized $^3$He can successfully be injected into the liquid
helium-filled collection volume in the nEDM experiment.  A successful injection critically depends on having a low
pressure inside the injection tube to minimize scattering of the $^3$He atoms from the $^4$He vapor.  Superfluid-$^4$He
film that creeps up from the collection volume into the injection tube and evaporates will increase the pressure inside
the tube.  Hence, a film burner was designed to control the superfluid-$^4$He film so that it does not vaporize into the
flight tube.  The $^4$He-film burner has been manufactured and is ready to be tested.  Most of the parts for the
injection test---including the 1.2~kG tri-coil magnet, the gas handling system, plus the ABS and pulsed NMR
system---have been built and, will be whenever possible, tested.  

The injection tests are presently delayed due to high voltage R\&D activities utilizing cryogenic components required
for these tests.  On a shorter timescale however, we have shifted focus to carry out a smaller scale experiment in
collaboration with the University of Illinois at Urbana-Champaign that will test just the film burner itself.  In this
film-burner test, no $^3$He will be injected, and thus less cooling power will be required so that a smaller dilution
refrigerator (DR) can be utilized.

\section{Neutron Beamline Design}

As part of the Fundamental Neutron Physics Beamline (FnPB) construction project at the SNS, TUNL has been in charge of
modeling the neutronics performance of both the polychromatic and the 0.89~nm beam lines.  As a natural extension to
this effort, in collaboration with the University of Kentucky, we are working to extend this work so as to benchmark the
flux measurements to the nEDM experiment in order to provide a more accurate estimate of the neutron fluence into the
measurement cells as well as optimize the design of the neutron guides.

The FnPB facility includes two neutron beamlines: a polychromatic cold beam and a 0.89~nm monochromatic beam.  For the
former, cold neutrons from the liquid hydrogen moderator are transported approximately 15~m through a series of
$10~\mbox{cm}\times 12$~cm straight and curved guides to the experimental area in the main SNS target building.  For the
latter, a 0.89~nm monochromatic beam is reflected out of the polychromatic beam by a stage-1, potassium-intercalated
graphite monochromator, followed by a second stage-1, rubidium-intercalated graphite monochromator configured in a
double-crystal design.  The resulting beam is almost parallel to the polychromatic beam and directed towards the
external nEDM building.  Measurements of the neutron flux were made at the end of the cold beamline and at the end of
the 8~m expansion guide on the 0.89~nm beamline\cite{Fom09}.

Simulations of the neutron transport to the nEDM measurement cells were performed using both the {\sc mcstas} and {\sc
geant4} codes\cite{Lef99,Ago03}.  The neutron flux was measured at the end of the cold guide in 2009\cite{Fom09}.
When compared with the two simulations which are consistent with each other, it indicates that the measured flux at
0.89~nm is roughly a factor of 20~\% higher than the flux predicted by simulation.  Similar comparisons were made to the
measured neutron flux at the end of the 8~m expansion horn on the 0.89~nm guide as well.  These simulations indicated
that the measured flux at 0.89~nm is less than the predicted value by a factor of roughly 2.4.  Based on the good
agreement with the cold beam, we postulated that this difference arises from unknown loss mechanisms within the
monochromator crystals.

The neutron flux to the nEDM measurement cells was simulated for each of the two beamlines.  For the polychromatic line,
the last 3.5~m of existing straight guide was removed and replaced by a ballistic guide/polarizing system extending to
the cells.  The new guide geometry consists of a $10 \times 12$~cm$^{2}$ polarizing bender, 4.25~m long, 22~m radius,
and 5~channels, followed by an 8~m long ballistic expansion horn that expands from the $10 \times 12$~cm$^{2}$ entrance
to a $30 \times 30$~cm$^{2}$ exit.  The expansion horn is followed by a $30 \times 30$~cm$^{2}$, $m=1.5$ straight guide,
14.15~m long that extends from the exit of the expansion horn to the entrance of the beam splitter.  The splitter is two
almost parallel, 6~m long converging guides that have a $15 \times 30$~cm$^{2}$ entrance and $7 \times 14$~cm$^{2}$ exit.
The midpoint of each measurement cell is 0.78~m downstream of the exit of the guide.

The neutron flux into the measurement cells was calculated using both codes, although polarization calculations in the
polarizing bender are presently only calculated using {\sc geant4} and these polarizations are used to scale the {\sc
mcstas} results.  Agreement between the two codes is of the order of a few percent.  At the nominal operating power of
2~MW, we estimate the 0.89~nm flux using the cold line to be
\begin{equation}
    \frac{d\phi_{cold}}{d\lambda} = 9.8 \times 10^{6}~\mbox{n/s/cm}^{2}\mbox{/\AA}
    \quad \mbox{and for the UCN line} \quad
    \frac{d\phi_{UCN}}{d\lambda} = 2.0 \times 10^{6}~\mbox{n/s/cm}^{2}\mbox{/\AA}.
\end{equation}
These fluxes correspond to UCN production rates in the measurement cells of 
\begin{equation}
    \frac{dN_{UCN}}{dt} = 1290~\mbox{UCN/s}
    \quad \mbox{and for the UCN line}\quad
    \frac{dN_{UCN}}{dt} = 263~\mbox{UCN/s}.
\end{equation}

One can see that the flux using the polychromatic line is a factor of five larger than the monochromatic line.  The 
collaboration has decided to utilize the polychromatic line for the nEDM measurements and thus additional optimization 
of this beamline is underway.  

\section{Cryogenic Design and Testing}

The neutron electric dipole moment (nEDM) experiment requires cooling a 1~m$^3$ volume of superfluid liquid helium to an
operating temperature between 0.25~K and 0.45~K. This central volume encloses a separate target volume of superfluid in
which 0.89~nm neutrons from the SNS are down-scattered into UCNs and confined in an acrylic measurement cell.  The
central volume and target, plus associated experimental services, are cooled by a high-flow $^3$He-$^4$He dilution
refrigerator.  The vacuum container enclosing the cryogenic systems is called the cryovessel and is cooled using liquid
helium through a permanent connection to a helium liquefier system.  TUNL is responsible for the design and construction
of the cryovessel with its enclosed 80~K and 4~K thermal shields, the helium liquefier system, the $^3$He-$^4$He
dilution refrigerator, the central volume, and the associated vacuum hardware, cryogenic sensors, and controls.

\mbox{}\newline\noindent\textbf{Helium Liquefier System}

A dedicated system with a helium liquefier and helium flow to the cryovessel has been adopted.  Our initial design
coupled a Linde L70 liquefier directly to the cryovessel.  Recently, we have developed a Memorandum of Understanding
with the SNS for the construction and operation of a larger common helium liquefier that will serve the SNS users, the
nEDM experiment, and another proposed SNS user project.  In this plan, the nEDM experiment will receive liquid helium
and return cold or warm helium gas to a liquefier located in a building adjacent to the FnPB external building.  This
arrangement should improve the reliability of helium supply to the nEDM cryovessel and reduce maintenance requirements.
The nEDM experiment will maintain a large liquid-helium (LHe) storage volume and piping system to connect to the
cryovessel.  The design of the flow system is driven by the need to fill the 1 m$^3$ central volume with liquid helium
at 4~K and to cool the rest of the cryovessel as quickly and efficiently as possible.  To reduce vibrations, the 80~K
thermal shields will be cooled by a positive pressure flow of liquid nitrogen driven by an external cryogenic pump and
phase separator.

\mbox{}\newline\noindent\textbf{$^\textbf{3}$He-$^\textbf{4}$He Dilution Refrigerator}

Cooling the central volume and target cell from 4~K to 0.25~K requires the use of a large $^3$He-$^4$He dilution
refrigerator (DR).  This DR must provide a cooling power of 80~mW at a mixing chamber temperature of 300~mK. As such,
the dilution refrigerator will be among the most powerful ever built.  

We have studied configurations of several high cooling power machines, many of them used for polarized nuclear target
systems.  We also contracted Janis Research to produce a design study for the cryostat, pumping stacks and gas-handling
system for the nEDM dilution refrigerator.  The final design is meant to operate at a $^3$He flow rate of 20 to
30~mmol/sec.  This high flow rate requires a gas heat exchanger -- a separator -- which uses helium vapor from the
entrainment volume to cool the incoming $^3$He.  There is also a 1.5~K liquid helium evaporator to liquefy the $^3$He
before it reaches the still of the dilution refrigerator.  

Working with ORNL, we have modified the design so that it can function in addition as a $^3$He or $^4$He evaporative
refrigerator above 1~K. This is achieved by adding a flow circuit to recirculate $^4$He and a ``stopper'' below the
still that can short-circuit the low-temperature heat exchangers and increase the effective pumping speed of the
recirculation system.  Final design of the dilution refrigerator is underway.

\mbox{}\newline\noindent\textbf{Cryovessel}

The cryovessel serves not only to provide the vacuum and 4~K cryogenic environment for the experiment, it is the primary
safety envelope for the entire experiment.  As such it must be an ASME certified vacuum vessel and meet the SNS safety
requirements.

The cryovessel consists of an outer vacuum enclosure with internal 80~K and 4~K thermal shields.  We have modeled the
heat inputs into the liquid nitrogen (LN) cooled shield and the 4~K helium-cooled shield of the cryovessel.  The shields
will be insulated by vacuum and multiple layers of aluminized mylar superinsulation.  The shields will be constructed of
aluminum alloy with multiple cooling tubes for LN or He gas welded to the surfaces.  Because of the overall heat load to
the cryovessel, an 80~K LN cooled shield was chosen in preference to a He gas-cooled intermediate shield.  The 4~K
shield will be cooled through a thermosyphon arrangement, where evaporating helium from the entrainment volume is cycled
through the shield and then to the cold return line of the helium liquefier.  In the design process, we have calculated
scenarios for cooling the cryovessel and contents to 4~K and the final operating temperatures for the nEDM experiment.
Because of the large masses of the components and the 1~m$^3$ volume of LHe in the insulation volume, a cooldown will
require approximately 3--4 weeks.

Because the cryovessel will contain about 1~m$^3$ of superfluid liquid helium it presents unusual safety challenges.  
In addition to the DOE requirement that the vessel be an ASME certified vacuum vessel, it must also provide the safety 
envelope for the experiment under accident scenarios where the vacuum is lost and the liquid helium is vaporized.  We
are working with personnel at the SNS to develop a suitable
safety plan, calculate the effects of possible pressure release scenarios, and design mitigating vent and pressure
release devices.  Although the cryovessel will be constructed in accordance with ASME Pressure Code Section VIII, the
neutron windows and other external connections, must be separately designed and approved.  The pressure releases must
assure that no gases are released toward the neutron guides and that the internal cryovessel does not exceed its maximum
allowable working pressure of about 1.5~atm.

\mbox{}\newline\noindent\textbf{Cryogenic Test Facilities}

In the nEDM experiment, tubes with diameters of 0.5~inch and larger will connect volumes of superfluid liquid helium
under vapor pressure at $T \leq 0.5$~K to volumes at temperatures above the superfluid transition temperature.  It is
well-known that there can be large heat flows due to the flow of superfluid film up the tube as well as to the reflux of
warm gas caused by the evaporation of the film at higher temperatures\cite{Nac94,Hay94}.  The details of the reflux
process have implications for the design of the $^4$He film burners and of the heat sinking of the tubes in the nEDM
apparatus.  Previous theoretical and experimental results on this process have described relatively thin tubes, much
smaller than those planned for the nEDM configuration.  Using calculations based on the Cornut model, we have prepared a
sample cell for measurements of the superfluid heat flows in tubes of various lengths and diameters.  This will be an
independent test of the reflux model extended to the temperature range of importance to the nEDM experiment.

We have constructed at TUNL a cryogen-free cryostat and vacuum system for the testing of seals and construction
materials to temperatures below the helium superfluid transition.  It will be possible to cool half-scale prototypes as
large as 21~inches in diameter and test them in vacuum.  A Sumitomo 1.5~W (at 4~K) RDK-415D Gifford-McMahon
cryocooler with an F-50 water-cooled compressor refrigerates the test volume.  A separate sealed helium-evaporator
insert into the test volume will continue the cooling to less than 2~K. The fixtures being tested will be filled with
superfluid helium at 1 atmosphere of pressure to mimic the nEDM operating conditions.  Initially we are testing the
cooling and sealing of G-10 composite flanges with o-rings made from 0.005~inch thick Kapton foils\cite{Edw97}.  These
will be followed by tests of sealing and diffusion through candidate materials for neutron windows.

\section{Data Acquisition/Slow Controls}

TUNL has taken responsibility for the design and implementation of both the fast waveform digitization system as well as
the slow controls for the nEDM project.

The nEDM scintillation signal as a function of time is a measure of the beat frequency between the neutron and $^{3}$He
precession rates.  The Data Acquisition (DAQ) system will be triggered by prompt coincidence signals from an array of
photomultiplier tubes which will be digitized using a series of transient waveform digitizers to capture both the prompt
and after-pulse scintillation signals for of order a microsecond after each trigger event.  Event rates are expected to
be of order 1~kHz.  As an initial effort in the design of this system, we have set up a 100~MHz VME digitizer and used it
to capture a sample signal consisting of a main pulse with afterpulses.  We plan additional characterization of this
system prior to the procurement of the 1~GHz system that is envisioned for the main apparatus.

The proposed slow control system is designed around the use of VME crates with VxWorks as the on-board operating system.
\textsc{epic}s (Experimental Physics and Industrial Controls System) software which is a standard network-based control
package that is used at many major accelerator facilities (including ORNL/SNS) will provide the interface between the
many commercially available devices and the VME system.  \textsc{epic}s is open source software that operates in a
distributed network environment and is available to users at no cost.

The nEDM apparatus and measurement cycles will require control or monitoring of order 1,000 parameters and status
values.  The present plan is to develop five individual subsystems that will be built and tested at different sites as
the components they control are being developed.  As a distributed-control environment, \textsc{epic}s is ideally suited
for this application because the individual subsystem development can proceed independently up until the integration
step at final assembly and testing at the SNS.

\section{Summary}

The wide range of research backgrounds, technical expertise and numbers of graduate students within the TUNL research
community enable and stimulate a broad collaborative research program that spans several frontier areas in nuclear
physics.  TUNL faculty strive to continue their emphasis on maintaining a research environment that is conducive to
graduate education by engaging students in all aspects of research from concept development through data interpretation
and dissemination of results.  Collaboration of the research groups from the consortium universities along with the
research infrastructure at TUNL enable TUNL groups to not only take significant responsibilities in projects such as the
nEDM, but also contribute in a wide range of areas.  In addition, the small-group environment at TUNL provides
opportunities for students to work closely with postdocs and faculty, to have hands-on research experiences, and
ultimately grow to provide leadership roles in their individual projects.



\end{document}